\documentclass[10pt, a4paper]{article} 

\usepackage{indentfirst}

\usepackage{xcolor}

\usepackage{ulem}
\usepackage{amsthm}
\usepackage{marginnote}

\usepackage{booktabs}
\usepackage{tabularx}
\usepackage{array}
\usepackage{makecell}
\usepackage{siunitx}

\newcolumntype{Y}{>{\raggedright\arraybackslash}X}

\usepackage[english]{babel}
\usepackage{graphicx} 

\usepackage{cite}

\usepackage[thinc]{esdiff}

\usepackage{tabularx}

\usepackage{latexsym}
\usepackage{amsfonts} 
\usepackage{amsmath} 

\usepackage{multirow}

\usepackage{enumitem}

\usepackage[utf8]{inputenc} 
\usepackage[T1,T2A]{fontenc}

\usepackage{layout}

\usepackage[left]{lineno}

\usepackage{color, soul} 
\setulcolor{yellow}

\usepackage[colorlinks=true, citecolor=blue]{hyperref}

\usepackage[nameinlink]{cleveref}

\newtheorem*{theorem*}{Theorem}

\newtheorem*{note*}{Remark}
\newtheorem*{def*}{Definition}

\title{The Revenue Effect of Demand Misspecification in Event Ticket Pricing}

\author{Lev Razumovskiy \and Nikolay Karenin \and Mikhail Safro}
\newcommand{\Addresses}{{
\bigskip
\footnotesize

Lev Razumovskiy	, \textsc{RAMAX Group}\par\nopagebreak
  \textit{E-mail address} : \text{lev.razumovskiy@ramax.com}
  
\medskip

Nikolay Karenin, \textsc{RAMAX Group}\par\nopagebreak
  \textit{E-mail address} : \text{nikolay.karenin@ramax.com}

\medskip
Mikhail Safro, \textsc{RAMAX Group}\par\nopagebreak
  \textit{E-mail address} : \text{msafro@ramax.com}

}}
\date{}

\begin{document}
\maketitle

\begin{abstract}
We study a finite-horizon dynamic pricing problem for event tickets with limited inventory and time-varying demand. The central practical difficulty is that the total demand function $L(t)$ is not observed directly and must be estimated from data, while pricing decisions are sensitive to its temporal shape. The paper examines how the accuracy of this estimate affects revenue.

We consider a model in which sales intensity is driven by the total demand $L(t)$, a price-response function $v(p)$, and a time-dependent willingness-to-pay factor $\varphi(t)$. The factor $\varphi(t)$ plays a central role: it captures the increase in customers' willingness to pay as the event date approaches and makes the temporal profile of demand economically important for pricing. Within this framework, the updated numerical study evaluates a benchmark dynamic-programming policy across nine deterministic true-demand scenarios, a collection of feature-aware misspecifications of $L(t)$, and multiple environment regimes induced by $v(p)=e^{-\eta p}$, the deadline factor $\varphi(t)$, and inventory level $Q$. The reported summaries are based on stochastic simulation and a ratio-of-means relative-loss metric.
The results show that a more accurate representation of the temporal demand profile leads to more effective pricing decisions and higher revenue. Over the full misspecification collection the aggregate relative revenue loss is $0.42\%$, the upper decile exceeds $1\%$, and the most expensive errors are omissions of late-demand components. The average effect is therefore modest but non-negligible, and it becomes stronger when deadline effects are pronounced and inventory is tight.
\end{abstract}

\section*{Introduction}

Dynamic pricing for event tickets is a standard problem in revenue management. In practice, most difficulties arise from the lack of reliable information about total demand, that is, about the intensity of potential customer arrivals and how it evolves over time. The same event may exhibit a long period of low baseline activity, a surge induced by marketing, and a pronounced pre-event spike, and these components can interact with pricing and inventory constraints in nontrivial ways. As a result, even well-established pricing rules can perform poorly when the temporal profile of total demand is mischaracterized.

This paper examines a specific question motivated by this observation: how does revenue change when pricing decisions are based on a more accurate estimate of the total demand profile rather than on a plausible but misspecified estimate? We address this question through a controlled experimental design that isolates the effect of the estimated total demand profile while keeping the remaining modeling assumptions fixed. In the updated numerical study, this comparison is carried out over nine deterministic true-demand scenarios, a feature-aware misspecification collection for $\widehat L(t)$, and multiple market environments generated by varying $\eta$, $\varphi(t)$, and inventory.

The present paper does not attempt to solve the estimation problem for $L(t)$ itself. Instead, it studies why the accuracy of its estimation matters economically: different approximations of $L(t)$ lead to different pricing policies and, consequently, to different revenue outcomes. To place this analysis in a practical context, Section~4 briefly discusses how $L(t)$ may be estimated in real ticketing systems using observable data and interpretable demand components, and how the experimental revenue effects should be interpreted in business terms.



\paragraph{Literature Overview.}

Dynamic pricing and revenue management constitute a broad research area at the intersection of operations research, management science, economics, and applied mathematics \cite{EK03,vdB15,ChenChen15,TVR04,Phillips21}. A central problem in this literature is the sale of finite inventory over a limited horizon in settings where unsold capacity becomes worthless after a fixed deadline and prices must be adjusted in response to both the remaining stock and the temporal structure of demand \cite{EK03,TVR04,Phillips21}.


The finite-horizon inventory-pricing problem originates in the early literature on optimal pricing under limited stock \cite{KincaidDarling63}. In a revenue-management setting, one of the most influential formulations is the stochastic finite-horizon model of Gallego and van Ryzin, which established a benchmark framework for the dynamic pricing of perishable assets \cite{GallegoVanRyzin94}. Subsequent work extended this framework in several practically important directions, including markdown and promotional pricing \cite{FengGallego95}, periodic pricing of seasonal products \cite{BitranMondschein97}, models with time-dependent and nonhomogeneous demand \cite{FengGallego00,ZhaoZheng00}, and multi-product or retail-chain settings \cite{GallegoVanRyzin97,BitranCaldenteyMondschein98}. A related stream of work studies pricing with intermediate sales or revenue targets; in particular, Besbes and Maglaras show that milestone constraints give rise to feedback-form pricing policies driven by downstream requirements \cite{BesbesMaglaras12}.


A separate and highly relevant strand of the literature studies pricing under incomplete demand information
\cite{vdB15,ChenChen15}. These works emphasize that, in many applications, the seller does not know the demand system in advance and must estimate it from transaction data while making pricing decisions \cite{vdB15,ChenChen15}. More recent research has also incorporated data-driven and machine-learning-based methods for learning customer preferences and adjusting prices dynamically in perishable-product environments \cite{YangChuWu22}. This direction is especially important for ticket sales, where the seller typically observes only realized purchases rather than the full latent market demand.

The event-ticket setting involves several mechanisms that are not always explicitly captured in standard airline or hotel models. Tickets are a canonical example of a perishable good, and their pricing is shaped by resale opportunities, intertemporal substitution, and the approaching event date \cite{Courty03,Sweeting12}. In the primary market, ticket pricing has been studied under demand uncertainty and posted-price rigidity \cite{JonesYeoman09}, in connection with strategic consumer behavior and inventory-information disclosure \cite{HuangWangHo17}, and in empirical studies in sport management that examine the determinants of dynamic ticket prices \cite{DrayerShapiroLee12,ShapiroDrayer14}. The literature also shows that demand-based ticket pricing is evaluated not only in terms of revenue outcomes but also through consumer perceptions of fairness and value \cite{ShapiroDrayerDwyer16}.


Against this background, the present paper sits at the intersection of classical finite-inventory dynamic pricing and applied ticket-sales modeling. In the revenue-management tradition, we consider a finite stock and a finite sales horizon \cite{GallegoVanRyzin94,TVR04}. Consistent with the literature on limited demand information, we focus on the effects of correctly specifying or misspecifying the aggregate demand profile \cite{vdB15,ChenChen15}. At the same time, our formulation is tailored to ticket sales, where the temporal pattern of demand is crucial because purchasing intensity may vary substantially over the selling horizon, particularly near the event date \cite{Courty03,Sweeting12}.

From this perspective, we model the purchase intensity as
$$
\lambda(t, p)=L(t)\,v\left(\frac{p}{\varphi(t)}\right),
$$
where $L(t)$ captures the temporal profile of total demand and $\varphi(t)$ is a time-dependent willingness-to-pay adjustment. The factor $\varphi(t)$ reflects the empirically observed increase in customers’ willingness to pay as the event date approaches and plays a crucial role in generating non-trivial pricing dynamics. Without the time-dependent factor $\varphi(t)$, the revenue differences across alternative demand profiles would be much weaker in our experiments, since the timing of demand would matter substantially less for pricing.

The chosen willingness-to-pay dynamics are intended to reflect a practical pattern often observed in concert-type events and similar sales settings, where purchasing activity rises sharply near the end of the selling horizon. In this sense, $\varphi(t)$ increases the importance of demand timing for pricing and makes differences between demand profiles more visible in the benchmark experiment.

This specification allows us to
isolate the effect of misspecifying the temporal demand profile itself. This mechanism is less explicit in much of the existing literature, which often focuses either on stochastic control or on online learning. Our approach is therefore to make explicit how the accuracy of the approximation to $L(t)$ affects the resulting pricing policy and the seller's revenue within a stochastic dynamic-programming benchmark.

We consider a ticket sales process that starts at time $t=0$ and must finish at time $t=T$.
Denote by $Q$ the total number of tickets to be sold over this period, and let $x(t)$ be the
remaining inventory. We assume that the total demand (customer flow) is a time-varying
function $L(t)$.
Let $p(t)$ be the ticket price at time $t$, and introduce a price-response function
$v(\cdot)$ that takes values in $[0, 1]$. The function $v\left(\frac{p(t)}{\varphi(t)}\right), t = 0, 1, \ldots, T$ represents the fraction of customer flow willing to purchase at time $t$ and price $p(t)$.

In the absence of inventory constraints, the mean sales rate and the mean revenue rate at time $t$ can be
written respectively as
$$
L(t)\,v\left(\frac{p(t)}{\varphi(t)}\right), \qquad p(t)\,L(t)\,v \left(\frac{p(t)}{\varphi(t)}\right).
$$
With finite inventory, these expressions remain valid as the unconstrained rates until the stock is exhausted; once sell-out is reached, actual sales are truncated by the remaining inventory, and no further revenue is generated thereafter.

The objective of the model is to determine a pricing policy $p(\cdot)$ that maximizes the total
expected revenue over the horizon $T$.

A central practical difficulty is that the profile $L(t)$ is not observed directly and must be
estimated from data. The main question studied in this paper is therefore the following:
\emph{what is the revenue impact of using a more accurate estimate of $L(t)$, compared to a
reasonable but misspecified estimate?}
To isolate this effect, we keep the remaining components of the pricing model fixed (including
the functional form of $v(\cdot)$ and the admissible price range) and vary only the information
about $L(t)$ used by the pricing strategy.

We therefore compare an \textit{oracle} regime, in which the policy is computed using the true scenario-specific $L(t)$, with a structured collection of \textit{feature-aware misspecified} regimes, in which the proxy $\widehat L(t)$ preserves the overall demand scale but distorts specific structural components such as peak timing, peak height, plateau level, smoothing, growth or decay slope, and late growth. The experiment covers nine deterministic true-demand scenarios and five misspecifications per scenario, so the collection contains $45$ non-oracle policies in addition to the oracle benchmark. For each scenario--misspecification pair we evaluate $27$ environments obtained by varying $\eta \in \{0.0075,0.0100,0.0130\}$, three deadline-effect regimes for $\varphi(t)$, and inventory levels $Q \in \{500,700,900\}$.

In all numerical experiments the sales process is simulated under the true $L(t)$ using common random numbers, so differences in revenue are attributable primarily to the quality of the total-demand profile used when constructing prices. Aggregated with the article's primary ratio-of-means relative-loss metric, the full misspecification collection yields a relative revenue loss of $0.42\%$ on average, with a $90$th percentile of $1.17\%$; the most expensive misspecifications are omissions of late-growth components ($2.26\%$) and omitted peaks ($0.77\%$).

The paper is organized as follows. Section~\ref{sec:basic_math_model} formulates the stochastic ticket-pricing model. Section~\ref{sec:pricing_policies} presents the dynamic-programming policy used as the numerical benchmark. Section~\ref{sec:experiments} reports the stochastic experiments across the scenario--misspecification--environment grid and summarizes the results both in absolute monetary terms and in relative loss in revenue. Section~\ref{sec:estimation} discusses limitations, practical approaches to estimating $L(t)$ from data, and the interpretation of the reported revenue effects.

\paragraph{Main results and contributions.}
The primary contribution of this paper is a controlled experimental framework that quantifies
the value of information about the total demand profile $L(t)$ in ticket pricing using nine deterministic demand scenarios, a collection of feature-aware misspecifications, and $27$ market environments. Secondly, the experiments identify which structural errors are most costly: omissions of late-growth components dominate the ranking with a $2.26\%$ ratio-of-means revenue loss, followed by omitted peaks at $0.77\%$, whereas timing and height perturbations are materially smaller. Finally, the study shows how this effect depends on the market regime: losses increase from $0.32\%$ under flat willingness-to-pay dynamics to $0.53\%$ under a strong deadline effect, and from $0.14\%$ at high inventory to $0.68\%$ at low inventory, while statistically significant oracle reversals are observed in only $0.41\%$ of scenario--environment combinations.

\section{Mathematical model of pricing for event tickets} \label{sec:basic_math_model}\label{sec:discrete_model}

Sales take place over a finite discrete-time horizon  $t = 0, 1,\ ...\ , T$, where $T$ denotes the final selling period before the event. Each period may correspond to a day or another fixed time interval.

At each time $t$, the seller sets a ticket price $p(t)$, which may vary over time. Prices are restricted to take values from a finite ordered set $\mathcal{P} = \{\pi_1 < \pi_2 <\ ...\ \pi_K\}$, reflecting practical constraints on admissible price levels. In particular, prices are bounded from below and above by $\pi_{\min}$ and $\pi_{\max}$. 

The seller has a finite inventory of tickets, denoted by $Q > 0$. Let $x(t)$ denote the remaining inventory at the beginning of period $t$, with $x(0) = Q$. Sales permanently reduce inventory, and no replenishment is allowed.

Demand in each period is stochastic and depends on both time and price. We introduce the following notation:

\begin{itemize}
    \item[] $L(t)$ -- a total demand function, capturing the overall level of interest in the event at time $t$;
    \item[] $p(t)$ -- a pricing policy;
    \item[] $v(p)$ -- a baseline price-response function, assumed to be decreasing in price;
    \item[] $\varphi(t)$ -- a time-dependent willingness-to-pay factor, reflecting the fact that customers become less price-sensitive as the event date approaches.
\end{itemize}

\paragraph{Remark (Demand estimation).}
Throughout this paper, the functions $v(p)$ and $\varphi(t)$ are treated as given and known.
The particular forms adopted here are intended to reflect a plausible and practically typical scenario, in line with both our experience and standard modeling choices in the literature.
The aggregate demand profile $L(t)$ is typically unobserved and must be inferred from external
signals (e.g., sales dynamics, engagement metrics, and marketing activity). In this work we focus
on the value of information about $L(t)$ for pricing: numerical experiments compare policies computed
under oracle access to $L(t)$ and under misspecified proxies for $L(t)$. Practical data sources and
estimation approaches for $L(t)$ are discussed in Section~\ref{sec:estimation}.

For a given price $p(t)$ at time $t$, the unconstrained demand is modeled as a Poisson random variable with intensity $\lambda(t, p(t)) := L(t) v\left(\frac{p(t)}{\varphi(t)}\right)$. 

Actual ticket purchases are limited by the remaining inventory and are given by 
$$
s(t) = \min(N(t), x(t)), \quad N(t) \sim \text{Pois}\Big(\lambda(t, p(t))\Big)
$$
Inventory evolves according to $x(t+1) = x(t) - s(t)$. This formulation ensures that sales cannot exceed the available inventory at any time.

Each ticket sold at time $t$ generates revenue equal to the current price $p(t)$.
The seller’s total revenue over the selling horizon is therefore a random variable given by

$$
R[p(\cdot)] := \displaystyle \sum^{T}_{t=0} p(t) s(t),
$$

that is, the cumulative revenue from all realized ticket purchases. The seller's objective is to choose a pricing policy $\{p(t)\}^{T}_{t=0}$ which maximizes the expected total revenue $\mathbb{E}R[p(\cdot)]$. 

We impose the following constraints on admissible pricing strategies:
\begin{itemize}
\item[1)] \textbf{Monotonicity constraint}: ticket prices are required to be non-decreasing over time~\footnote{This assumption is motivated by common industry practice and by our discussions with event organizers, who emphasized that price reductions after earlier purchases at higher prices may lead either to refund claims or, more broadly, to a loss of customer trust, making such pricing actions undesirable in practice.}, 
$$
p(t+1) \ge p(t).
$$
\item[2)] \textbf{Discrete pricing constraint}: prices must belong to the finite price grid $\mathcal{P}$.
\item[3)] \textbf{Inventory constraint}: sales in each period cannot exceed the remaining inventory.
\end{itemize}

The resulting problem is a finite-horizon stochastic dynamic pricing problem with remaining inventory $x(t)$
and price $p(t)$
as the control. 
The problem naturally admits a dynamic programming formulation, which we use as a benchmark in subsequent numerical experiments.
The presence of the factor $\varphi(t)$ plays a crucial role in generating non-trivial optimal pricing dynamics. Without a time-dependent willingness-to-pay component, the optimal price trajectory often degenerates to an approximately constant price. Introducing $\varphi(t)$ allows the model to capture empirically observed price increases as the event date approaches.

\section{Pricing policy} \label{sec:pricing_policies}



A pricing policy is a sequence of decision rules that selects a price level at each period based on the currently available information. Under the monotonicity constraint $p(t+1)\ge p(t)$ and the discrete price grid $\mathcal{P}=\{\pi_1<\cdots<\pi_K\}$, it is 
necessary
to augment the state by the index of the last posted price.

Let $V(t, x,i)$ denote the optimal expected revenue collected from periods $t,\dots, T$ given the remaining inventory $x \in \{0, \dots, Q\}$ and current price level $\pi_i$ (so that the admissible actions at time $t$ are $\{\pi_k \mid \,k\ge i \}$). Then $V(T+1, \cdot,\cdot)=0$ and the value function satisfies the Bellman recursion
\begin{equation}\label{eq:bellman_discrete_monotone}
V(t, x, i) = \max_{k\in\{i,\dots,K\}} \mathbb{E}\Big[ \pi_k\,\min\!\big(N_k(t),x\big) + V\!\Big(t+1, x-\min\!\big(N_k(t),x\big),\,k\Big)
\Big],
\end{equation}
where $N_k(t)\sim \text{Pois}\big(\lambda(t, \pi_k)\big)$ and $\lambda(t, p)=L(t)\,v \big(\frac{p}{\varphi(t)}\big)$. Any maximizer $k_t^*(x,i)$ in~\eqref{eq:bellman_discrete_monotone} defines an optimal feedback policy via $p(t)=\pi_{k_t^*(x(t),i_t)}$ and $i_{t+1}=k_t^*(x(t),i_t)$.


\section{Numerical experiments}\label{sec:experiments}

We now report numerical results for the pricing policy introduced above and compare its revenue performance under oracle and misspecified demand specifications. In all experiments, the sales process is simulated under one of the true demand profiles $L(t)$; only the demand profile used to compute the pricing policy is changed. This design isolates the effect of misspecifying $L(t)$ while keeping the rest of the setup fixed.

\begin{figure}[ht]
    \centering
    \makebox[\textwidth][c]{%
        \includegraphics[width=0.96\linewidth]{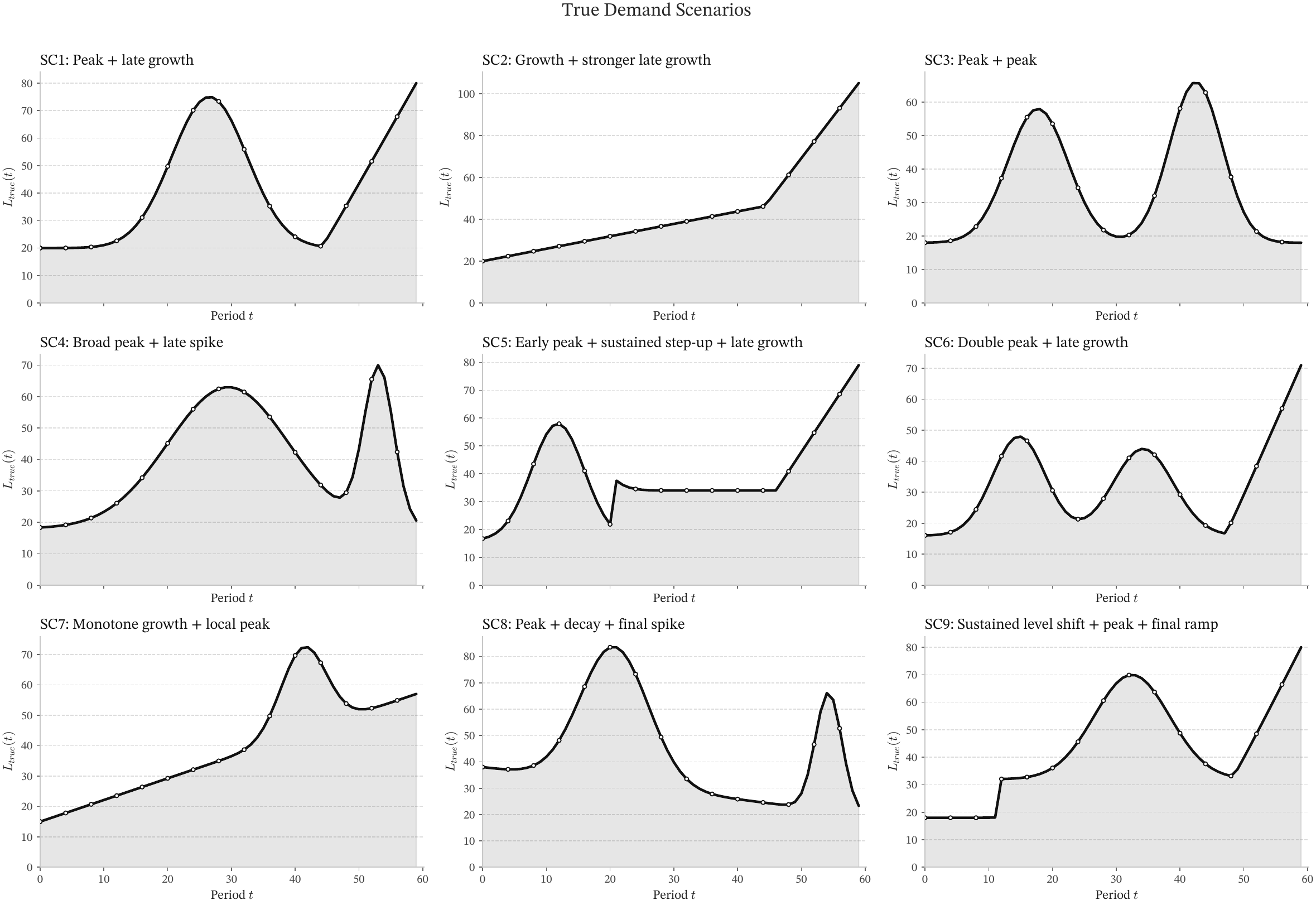}
    }
    \caption{Nine deterministic true-demand scenarios used in the experiment.}
    \label{fig:scenarios_true}
\end{figure}

\begin{figure}[ht]
    \centering
    \makebox[\textwidth][c]{%
        \includegraphics[width=0.96\linewidth]{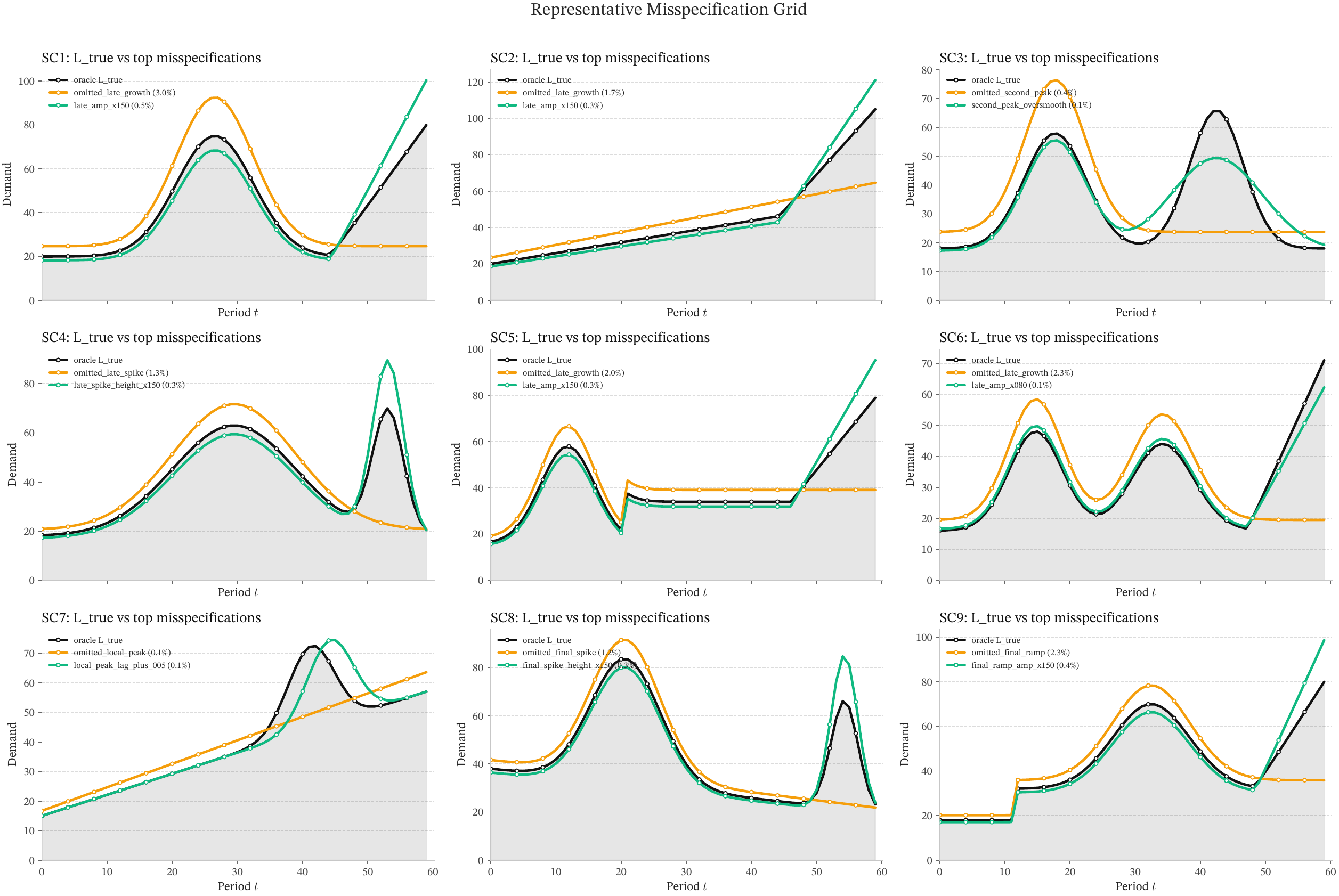}
    }
    \caption{Representative feature-aware misspecifications of $\widehat L(t)$ against the corresponding true scenarios.}
    \label{fig:misspec_grid}
\end{figure}

\begin{figure}[ht]
    \centering
    \makebox[\textwidth][c]{%
        \includegraphics[width=0.78\linewidth]{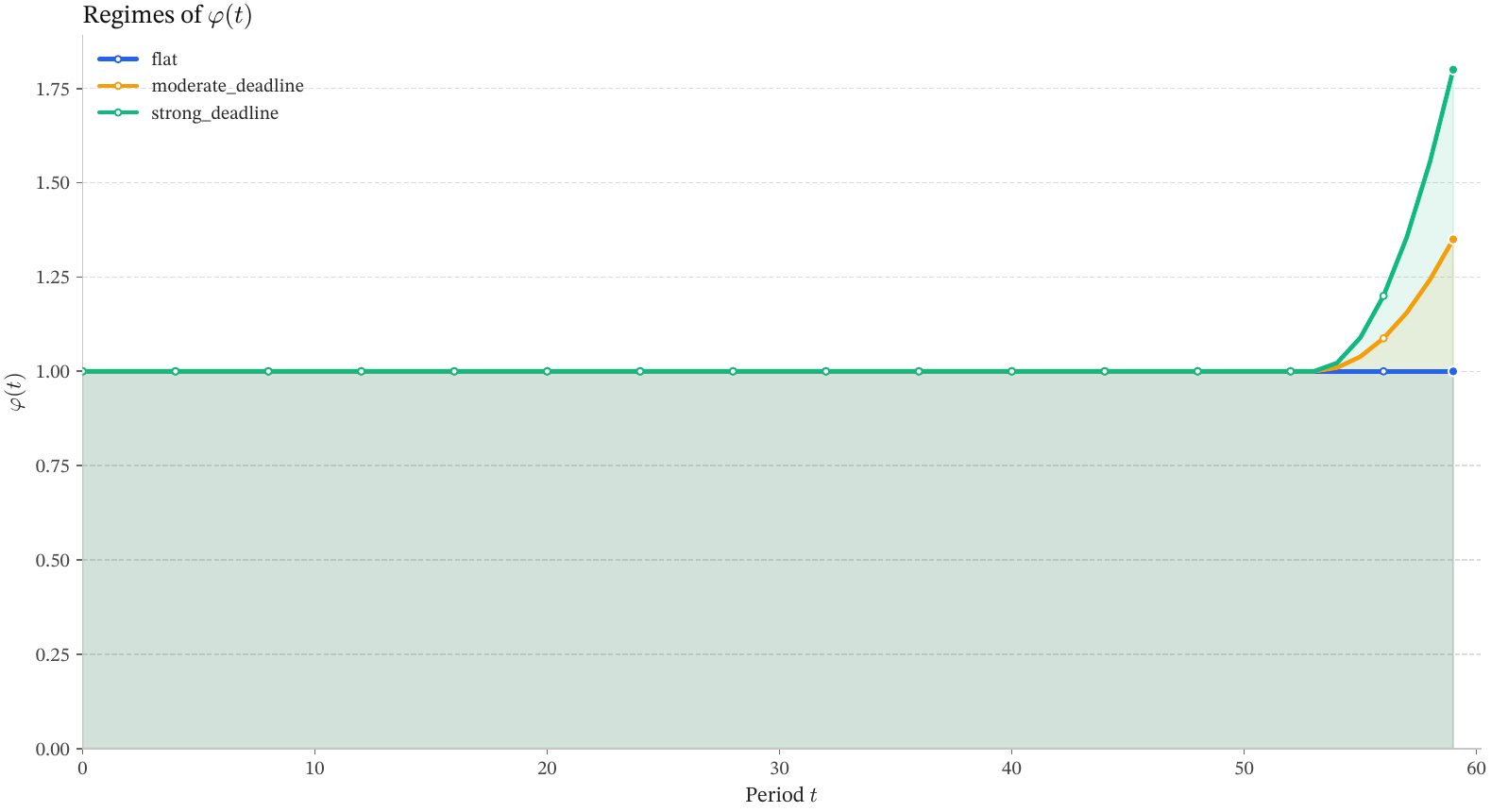}
    }
    \caption{Regimes of $\varphi(t)$.}
    \label{fig:phi_regimes}
\end{figure}

\begin{figure}[ht]
    \centering
    \makebox[\textwidth][c]{%
        \includegraphics[width=0.78\linewidth]{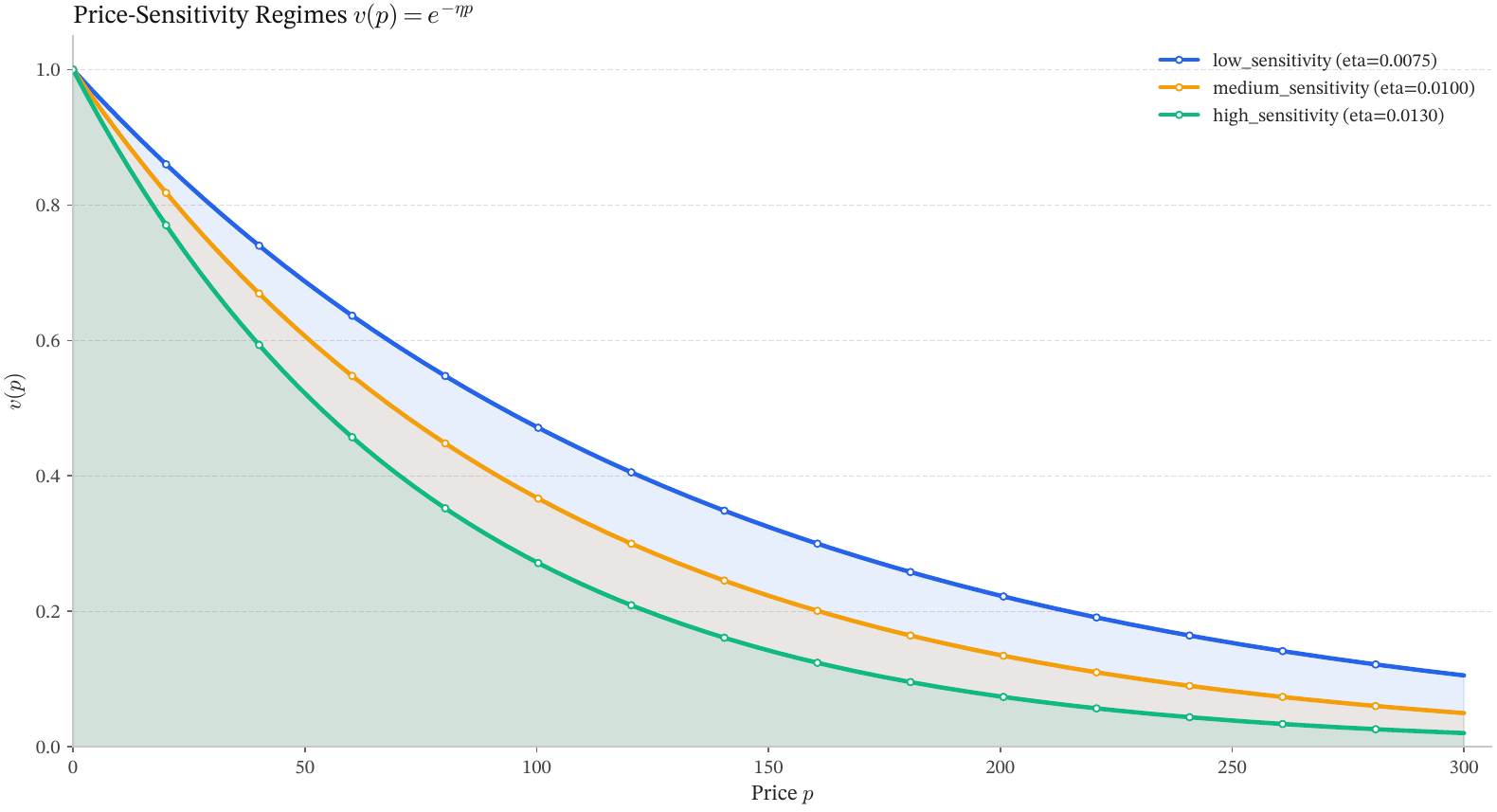}
    }
    \caption{Regimes of $v(p)=e^{-\eta p}$.}
    \label{fig:eta_regimes}
\end{figure}

\paragraph{Experimental design.}
The numerical experiment varies along three groups of parameters:

\begin{itemize}
    \item[-] \textit{True-demand scenarios.} 
    We consider nine deterministic true-demand scenarios for $L(t)$, shown in Figure~\ref{fig:scenarios_true}. Together, they span a range of temporal demand patterns, including isolated and repeated peaks, monotone growth, broad peaks followed by late spikes, sustained level shifts, post-peak decay, and final ramps. 

    \item[-] \textit{Feature-aware misspecifications.} 
    For each true scenario, we construct five misspecified proxies $\widehat L(t)$ by perturbing one or two structural components while leaving the rest of the demand profile unchanged. Depending on the scenario, the misspecification collection includes peak-timing errors, peak-height errors, omitted peaks, oversmoothing of narrow spikes, plateau-level errors, growth- or decay-slope errors, and late-growth timing or magnitude errors. Each non-oracle proxy is normalized to have the same total mass as its corresponding true scenario. Thus, before introducing the environment regimes, the design contains $45$ non-oracle scenario--proxy cases in addition to $9$ oracle cases. Representative examples are shown in Figure~\ref{fig:misspec_grid}.

    \item[-] \textit{Environment regimes and stochastic simulation.} We then introduce the environment layer of the experimental design by combining three price-sensitivity regimes, three deadline-effect regimes, and three inventory regimes:
    \[
    \eta \in \{0.0075,\,0.0100,\,0.0130\}, \qquad
    Q \in \{500,\,700,\,900\},
    \]
    The willingness-to-pay factor $\varphi(t)$ is specified in three forms: flat, moderately increasing near the deadline, and strongly increasing near the deadline (Figure~\ref{fig:phi_regimes}). The corresponding price-response curves $v(p)=e^{-\eta p}$ are shown in Figure~\ref{fig:eta_regimes}. This yields $27$ environments in total. For each scenario--misspecification--environment triple, performance is evaluated over $3000$ stochastic simulation runs with common random numbers.
\end{itemize}

To summarize the revenue effect of misspecification, we report both absolute revenue losses and the ratio-of-means relative loss
\[
\mathrm{RL}_{\mathrm{ROM}}
=
\frac{\sum_{m=1}^{M} R^{\mathrm{oracle}}_m-\sum_{m=1}^{M} R^{\mathrm{misspec}}_m}
{\sum_{m=1}^{M} R^{\mathrm{oracle}}_m},
\]
where $R^{\mathrm{oracle}}_m$ and $R^{\mathrm{misspec}}_m$ denote the revenues obtained in simulation run $m$ under the oracle and misspecified policies, respectively. The absolute loss provides the effect in monetary terms, whereas $\mathrm{RL}_{\mathrm{ROM}}$ is used as the primary scale-free metric for comparisons across cases.

\paragraph{Global revenue effect.}
Across the $1215$ non-oracle scenario--misspecification--environment cases, the mean oracle revenue is $88\,546.56$, whereas the mean revenue under the corresponding misspecified policies is $88\,175.58$. The implied mean absolute revenue loss is $370.98$ monetary units, and the aggregate ratio-of-means relative loss is $0.42\%$; the $90$th percentiles of the case-level loss distributions are $1\,010.21$ in monetary terms and $1.17\%$ in relative terms (Table~\ref{tab:global_summary}). 
The aggregate value of $0.42\%$ is averaged over the full set of misspecification cases, including many low-impact cases. When the comparison is restricted to the two most damaging misspecifications in each scenario, the mean relative revenue loss rises to approximately $0.94\%$.

\begin{table}[ht]
\centering
\small
\caption{Global summary over all non-oracle cases}
\label{tab:global_summary}
\begin{tabular}{l
S[table-format=4.0]
S[table-format=6.2]
S[table-format=6.2]
S[table-format=4.2]
S[table-format=4.2]
S[table-format=1.2]
S[table-format=1.2]}
\toprule
\makecell[l]{Cases} &
\multicolumn{1}{c}{\makecell{Mean oracle\\revenue}} &

\multicolumn{1}{c}{\makecell{Mean misspecified\\revenue}} &
\multicolumn{1}{c}{\makecell{Mean abs.\\loss}} &
\multicolumn{1}{c}{\makecell{Median abs.\\loss}} &
\multicolumn{1}{c}{\makecell{ROM rel.\\loss, \%}} &
\multicolumn{1}{c}{\makecell{$90$th pct. ROM\\rel. loss, \%}} \\
\midrule
1215 & 88546.56 & 88175.58 & 370.98 & 67.75 & 0.42 & 1.17 \\
\bottomrule
\end{tabular}
\end{table}

\paragraph{Which scenarios and error types are most dangerous?}
The largest mean losses are observed for SC1 (Peak + late growth, $0.80\%$), SC9 (Sustained level shift + peak + final ramp, $0.57\%$), and SC6 (Double peak + late growth, $0.51\%$). By contrast, the losses are much smaller for SC3 (Peak + peak, $0.14\%$) and especially SC7 (Monotone growth + local peak, $0.04\%$). The pattern becomes even clearer when the cases are grouped by error type: omitted late-growth components are the most costly misspecifications, with a ratio-of-means relative loss of $2.26\%$, followed by omitted peaks at $0.77\%$. Errors in late-growth magnitude remain non-negligible ($0.27\%$), whereas peak-timing and peak-height errors are substantially milder ($0.15\%$ and $0.13\%$). This ranking is conditional on the misspecification cases considered in the benchmark and should therefore be interpreted as a within-benchmark severity ordering rather than as a general ranking valid in all settings.

\begin{table}[ht]
\centering
\small
\caption{Error-type ranking by the primary article metric}
\label{tab:error_ranking}
\begin{tabular}{l
S[table-format=3.0,round-precision=0]
S[table-format=4.2]
S[table-format=1.2]}
\toprule
\makecell[l]{Error type} &
\multicolumn{1}{c}{Cases} &
\multicolumn{1}{c}{\makecell{Mean abs.\\loss}} &
\multicolumn{1}{c}{\makecell{ROM rel.\\loss, \%}} \\
\midrule
Omitted late growth & 135 & 2030.18 & 2.26 \\
Omitted peak & 108 & 670.56 & 0.77 \\
Late-growth magnitude error & 162 & 249.29 & 0.27 \\
Peak timing error & 243 & 127.00 & 0.15 \\
Peak height error & 189 & 108.25 & 0.13 \\
\bottomrule
\end{tabular}
\end{table}

\begin{figure}[ht]
    \centering
    \makebox[\textwidth][c]{%
        \includegraphics[width=0.98\linewidth]{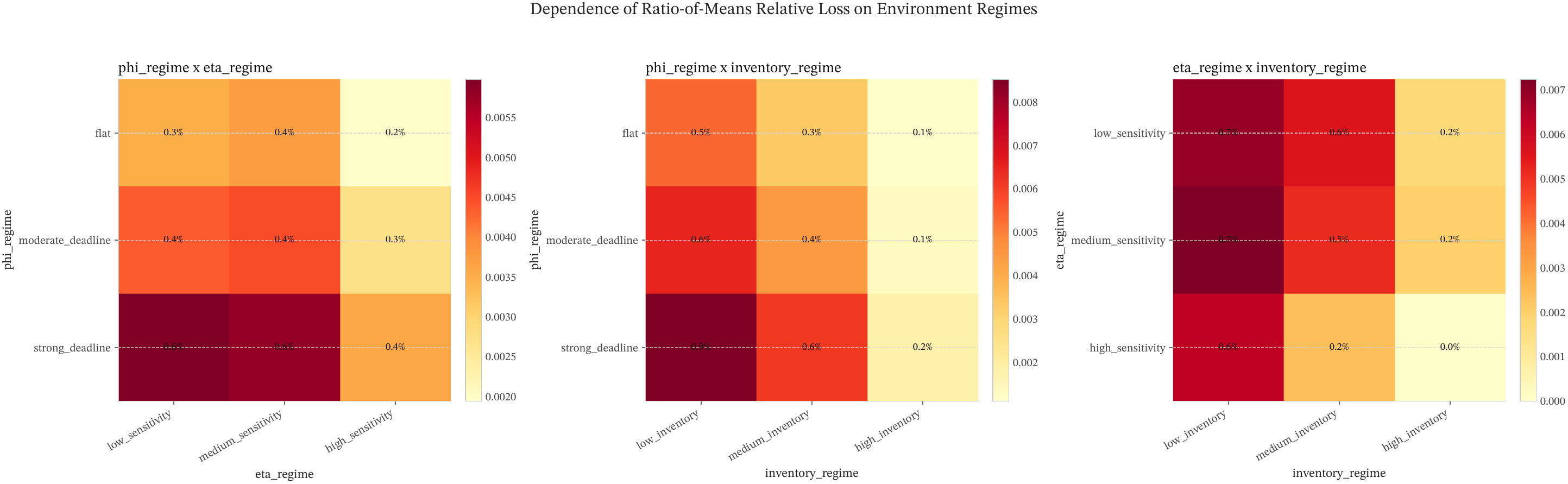}
    }
    \caption{Dependence of the ratio-of-means relative loss on the environment regimes.}
    \label{fig:env_heatmap}
\end{figure}

\paragraph{Dependence on market regimes.}
The losses increase systematically as the deadline effect becomes stronger: the mean ratio-of-means relative loss rises from $0.32\%$ under a flat $\varphi(t)$ regime to $0.40\%$ under a moderate deadline effect and to $0.53\%$ under a strong deadline effect. The losses are also strongly inventory-dependent: they equal $0.68\%$ at low inventory, $0.47\%$ at medium inventory, and only $0.14\%$ at high inventory. By contrast, higher price sensitivity attenuates the losses: the corresponding values are $0.46\%$, $0.47\%$, and $0.28\%$ for low-, medium-, and high-sensitivity environments, respectively. The largest losses occur in low-inventory, strong-deadline environments, where they reach $0.78\%$--$0.88\%$, whereas the smallest losses are observed in high-sensitivity, high-inventory environments, where they are zero in the benchmark grid considered here. These interactions are summarized in Figure~\ref{fig:env_heatmap}.

\begin{figure}[ht]
    \centering
    \makebox[\textwidth][c]{%
        \includegraphics[width=0.96\linewidth]{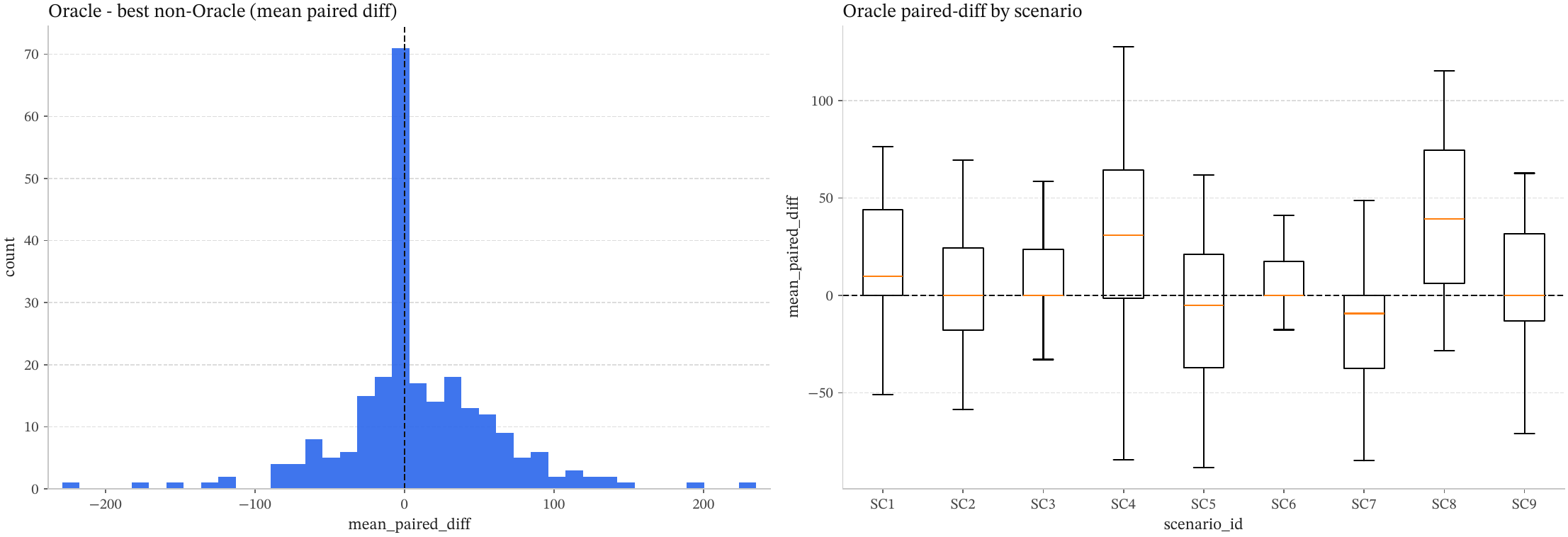}
    }
    \caption{Oracle benchmark check based on paired differences between the oracle policy and the best distinct non-oracle policy in each scenario--environment combination.}
    \label{fig:oracle_sanity}
\end{figure}

\paragraph{Oracle benchmark check.}
A useful diagnostic is whether the oracle policy is ever outperformed by a misspecified policy in stochastic simulation. For each scenario--environment combination, we compare the oracle policy with the best distinct non-oracle policy using paired revenue differences computed under common random numbers. A reversal is classified as statistically significant if the $95\%$ confidence interval for the mean paired difference $(R^{\mathrm{oracle}} - R^{\mathrm{best\ non\mbox{-}oracle}})$ lies entirely below zero. Across the $243$ scenario--environment combinations, the oracle policy has lower mean simulated revenue than the best distinct non-oracle policy in $34.16\%$ of raw comparisons, but statistically significant reversals occur in only $0.41\%$ of cases. The mean paired difference equals $9.09$, and the minimum observed value is $-228.94$. Thus, the apparent reversals are overwhelmingly small and are best interpreted as simulation noise rather than as evidence of meaningful oracle underperformance. Figure~\ref{fig:oracle_sanity} visualizes these paired differences.

\clearpage

\section{Practical remarks on estimating $L(t)$ from data}\label{sec:estimation}

The experiments in this paper are designed to isolate the value of information about the total demand profile $L(t)$.
In practice, $L(t)$ is latent and must be inferred from data. This section outlines the types of observable signals that can
support estimation of $L(t)$ and a practical decomposition of $L(t)$ into interpretable components.

At the same time, the numerical results in Section~\ref{sec:experiments} should be interpreted as a controlled sensitivity study. We keep $v(\cdot)$ and $\varphi(t)$ fixed and vary only the demand profile used to construct prices, so the reported losses quantify the revenue value of better information about $L(t)$ within the model rather than the full end-to-end forecasting error of a deployed ticketing system.

Typical data sources in ticketing systems include:
\begin{itemize}[leftmargin=2em]
  \item[-] \textbf{Sales trajectories:} time-stamped purchases, cancellations/returns (if applicable), and remaining inventory;
  \item[-] \textbf{Pricing and promotions:} posted prices, discount rules, promo codes, bundles, and other interventions~\footnote{In the present paper, however, the pricing model is restricted to non-decreasing posted prices, reflecting the practical concerns discussed above; accordingly, such interventions should be understood here as part of the broader empirical environment rather than as decision variables in the optimization model.};
  \item[-] \textbf{Engagement signals:} page views, sessions, add-to-cart events, checkout initiations, and conversion funnels;
  \item[-] \textbf{Marketing activity:} campaign schedules, budgets, impressions, clicks, reach, and attribution tags (e.g., UTM);
  \item[-] \textbf{Context and calendar features:} time-to-event, day-of-week, seasonality, venue capacity, artist lineup, and competing events;
  \item[-] \textbf{External attention proxies:} search trends, social media indicators, mailing-list activity, and press/news events.
\end{itemize}
These signals do not reveal $L(t)$ directly; rather, they provide partial information about audience interest and the intensity of exposure
over time.

A convenient representation for interpretation and modeling is
$$
L(t)=L_{\mathrm{base}}(t)+L_{\mathrm{ads}}(t)+L_{\mathrm{spike}}(t),
$$
where $L_{\mathrm{base}}(t)$ captures organic baseline interest and seasonal patterns, $L_{\mathrm{ads}}(t)$ captures the incremental customer flow induced
by marketing, and $L_{\mathrm{spike}}(t)$ captures a pre-event surge often observed close to the event date.

This decomposition is not unique, but it matches common operational intuition (baseline demand, campaign-driven uplift, and a deadline-driven spike)
and helps to interpret what a ``misspecification'' of $L(t)$ means in practice (e.g., omitting $L_{\mathrm{spike}}(t)$ or mis-timing the peak).


\paragraph{Baseline demand.}
One may estimate $L_{\mathrm{base}}(t)$ from historical events using regression or semi-parametric models on time-to-event and calendar covariates
(e.g., splines or generalized additive models). When multiple comparable events are available, hierarchical or mixed-effects formulations can
borrow strength across events while allowing event-specific heterogeneity.

\paragraph{Advertising component.}
Marketing effects can be captured by relating $L_{\mathrm{ads}}(t)$ to advertising exposures using lagged/adstock features. A standard pattern is that
advertising has delayed and decaying effects; this can be modeled with distributed lags or convolutional kernels on impressions/clicks. 
It is often difficult to separate the effect of advertising from the natural evolution of demand, especially when campaigns are launched in response to expected sales patterns. If advertising activity is scheduled independently of short-run demand fluctuations, its contribution to demand can be estimated more reliably.

\paragraph{Pre-event spike.}
The surge close to the event can be represented by a parametric term driven by the time remaining until $T$ (e.g., a convex increasing function on
$[T-\Delta,T]$) or by a state-space component capturing abrupt changes in intensity. In practice, failing to model this component is a common source
of systematic bias and motivates our ``Misspecified'' regime in numerical experiments.

\paragraph{Latent-intensity smoothing.}
Even when a parametric structure is specified, the raw estimate of $L(t)$ can be noisy. State-space models (e.g., local level/trend models) or
Bayesian filtering/smoothing can stabilize the inferred intensity while maintaining responsiveness to sudden changes.

\paragraph{Validation via backtesting.}
Since $L(t)$ is not observed, evaluation typically proceeds via predictive validation: does the inferred demand profile improve forecasts of sales
or conversions out-of-sample? Such backtesting is also the natural way to define what ``good'' means for a particular application.

\section*{Conclusions}

The paper examined the revenue effect of specifying the total demand function $L(t)$ with different degrees of accuracy in a finite-horizon ticket-pricing problem. In the updated experiment, this question was studied over nine deterministic true-demand scenarios, a collection of forty-five feature-aware misspecifications, and twenty-seven market environments generated by varying price sensitivity, deadline effects, and inventory. The numerical results showed that more accurate information about the total demand profile generally led to better pricing decisions and higher revenue in the stochastic dynamic-programming benchmark.

Measured by the article's primary ratio-of-means relative-loss metric, the aggregate loss over the full misspecification collection is $0.42\%$, with a $90$th percentile of $1.17\%$. The most expensive errors are structural omissions: omitted late-growth components produce a loss of $2.26\%$, and omitted peaks produce $0.77\%$. At the scenario level, the largest average losses are observed for Peak + late growth (SC1, $0.80\%$), Sustained level shift + peak + final ramp (SC9, $0.57\%$), and Double peak + late growth (SC6, $0.51\%$), while Peak + peak (SC3) and Monotone growth + local peak (SC7) are much more robust.

The environment summaries show that the value of correctly specifying $L(t)$ is not uniform across markets. The loss increases from $0.32\%$ under flat willingness-to-pay dynamics to $0.53\%$ under a strong deadline effect, and from $0.14\%$ at high inventory to $0.68\%$ at low inventory. It is smaller in high-sensitivity price regimes ($0.28\%$) than in low- or medium-sensitivity regimes ($0.46\%$ and $0.47\%$). The worst cases in the grid are therefore low-inventory, strong-deadline environments, whereas high-inventory high-sensitivity environments are the most robust.

The oracle sanity check further supports this interpretation. Although raw oracle reversals appear in $34.16\%$ of scenario--environment comparisons, statistically significant reversals occur in only $0.41\%$ of them. Hence the observed non-oracle wins are overwhelmingly attributable to simulation noise or numerically negligible differences rather than to substantive oracle failure.

\paragraph{Further directions.}

\begin{itemize}

    \item[-] \textit{Advertising-aware demand modeling} incorporated into the total-demand dynamics could improve the interpretability of the ticket-pricing model and help separate baseline demand from promotion-driven spikes and surges.

    \item[-] \textit{Broader experimental validation} on real or synthetic datasets. While the present paper already studies a broad synthetic benchmark with nine scenario families and multiple market environments, further work may place greater emphasis on data-driven experiments and on evaluating the proposed pricing policies across real events and richer empirical demand patterns.

    \item[-] \textit{Calibration and estimation} of $L(t)$ from operational signals (sales trajectories, engagement, and marketing activity) should be validated through systematic backtesting, to identify which temporal features of demand must be captured most accurately for reliable pricing decisions.
\end{itemize}

\bibliography{references}

@article{EK03,
  author  = {Elmaghraby, Wedad and Keskinocak, P{\i}nar},
  title   = {Dynamic Pricing in the Presence of Inventory Considerations: Research Overview, Current Practices, and Future Directions},
  journal = {Management Science},
  year    = {2003},
  volume  = {49},
  number  = {10},
  pages   = {1287--1309},
  doi     = {10.1287/mnsc.49.10.1287.17315}
}

@article{vdB15,
  author  = {den Boer, Arnoud V.},
  title   = {Dynamic Pricing and Learning: Historical Origins, Current Research, and New Directions},
  journal = {Surveys in Operations Research and Management Science},
  year    = {2015},
  volume  = {20},
  number  = {1},
  pages   = {1--18},
  doi     = {10.1016/j.sorms.2015.03.001}
}

@article{ChenChen15,
  author  = {Chen, Ming and Chen, Zhi-Long},
  title   = {Recent Developments in Dynamic Pricing Research: Multiple Products, Competition, and Limited Demand Information},
  journal = {Production and Operations Management},
  year    = {2015},
  volume  = {24},
  number  = {5},
  pages   = {704--731},
  doi     = {10.1111/poms.12295}
}

@book{TVR04,
  author    = {Talluri, Kalyan T. and van Ryzin, Garrett J.},
  title     = {The Theory and Practice of Revenue Management},
  publisher = {Springer},
  address   = {New York},
  year      = {2004},
  doi       = {10.1007/b139000}
}

@book{Phillips21,
  author    = {Phillips, Robert L.},
  title     = {Pricing and Revenue Optimization},
  edition   = {2},
  publisher = {Stanford University Press},
  address   = {Stanford, CA},
  year      = {2021}
}

@article{KincaidDarling63,
  author  = {Kincaid, W. M. and Darling, D. A.},
  title   = {An Inventory Pricing Problem},
  journal = {Journal of Mathematical Analysis and Applications},
  year    = {1963},
  volume  = {7},
  number  = {2},
  pages   = {183--208},
  doi     = {10.1016/0022-247X(63)90047-7}
}

@article{GallegoVanRyzin94,
  author  = {Gallego, Guillermo and van Ryzin, Garrett},
  title   = {Optimal Dynamic Pricing of Inventories with Stochastic Demand over Finite Horizons},
  journal = {Management Science},
  year    = {1994},
  volume  = {40},
  number  = {8},
  pages   = {999--1020},
  doi     = {10.1287/mnsc.40.8.999}
}

@article{FengGallego95,
  author  = {Feng, Youyi and Gallego, Guillermo},
  title   = {Optimal Starting Times for End-of-Season Sales and Optimal Stopping Times for Promotional Fares},
  journal = {Management Science},
  year    = {1995},
  volume  = {41},
  number  = {8},
  pages   = {1371--1391},
  doi     = {10.1287/mnsc.41.8.1371}
}

@article{BitranMondschein97,
  author  = {Bitran, Gabriel R. and Mondschein, Susana V.},
  title   = {Periodic Pricing of Seasonal Products in Retailing},
  journal = {Management Science},
  year    = {1997},
  volume  = {43},
  number  = {1},
  pages   = {64--79},
  doi     = {10.1287/mnsc.43.1.64}
}

@article{FengGallego00,
  author  = {Feng, Youyi and Gallego, Guillermo},
  title   = {Perishable Asset Revenue Management with Markovian Time Dependent Demand Intensities},
  journal = {Management Science},
  year    = {2000},
  volume  = {46},
  number  = {7},
  pages   = {941--956},
  doi     = {10.1287/mnsc.46.7.941.12035}
}

@article{ZhaoZheng00,
  author  = {Zhao, Wen and Zheng, Yu-Sheng},
  title   = {Optimal Dynamic Pricing for Perishable Assets with Nonhomogeneous Demand},
  journal = {Management Science},
  year    = {2000},
  volume  = {46},
  number  = {3},
  pages   = {375--388},
  doi     = {10.1287/mnsc.46.3.375.12063}
}

@article{BitranCaldenteyMondschein98,
  author  = {Bitran, Gabriel R. and Caldentey, Ren{\'e} and Mondschein, Susana V.},
  title   = {Coordinating Clearance Markdown Sales of Seasonal Products in Retail Chains},
  journal = {Operations Research},
  year    = {1998},
  volume  = {46},
  number  = {5},
  pages   = {609--624},
  doi     = {10.1287/opre.46.5.609}
}

@article{GallegoVanRyzin97,
  author  = {Gallego, Guillermo and van Ryzin, Garrett},
  title   = {A Multiproduct Dynamic Pricing Problem and Its Applications to Network Yield Management},
  journal = {Operations Research},
  year    = {1997},
  volume  = {45},
  number  = {1},
  pages   = {24--41},
  doi     = {10.1287/opre.45.1.24}
}

@article{BesbesMaglaras12,
  author  = {Besbes, Omar and Maglaras, Costis},
  title   = {Dynamic Pricing with Financial Milestones: Feedback-Form Policies},
  journal = {Management Science},
  year    = {2012},
  volume  = {58},
  number  = {9},
  pages   = {1715--1731},
  doi     = {10.1287/mnsc.1110.1513}
}

@article{Courty03,
  author  = {Courty, Pascal},
  title   = {Some Economics of Ticket Resale},
  journal = {Journal of Economic Perspectives},
  year    = {2003},
  volume  = {17},
  number  = {2},
  pages   = {85--97},
  doi     = {10.1257/089533003765888449}
}

@article{Sweeting12,
  author  = {Sweeting, Andrew},
  title   = {Dynamic Pricing Behavior in Perishable Goods Markets: Evidence from Secondary Markets for Major League Baseball Tickets},
  journal = {Journal of Political Economy},
  year    = {2012},
  volume  = {120},
  number  = {6},
  pages   = {1133--1172},
  doi     = {10.1086/669254}
}

@article{JonesYeoman09,
  author  = {Jones, Steven L. and Yeoman, John C.},
  title   = {The Promoter's Role in Ticket Pricing: Implications of Real Options for Optimal Posted Prices and Rationing},
  journal = {Journal of Business Research},
  year    = {2009},
  volume  = {62},
  number  = {11},
  pages   = {1187--1192},
  doi     = {10.1016/j.jbusres.2008.09.002}
}

@article{DrayerShapiroLee12,
  author  = {Drayer, Joris and Shapiro, Stephen L. and Lee, Seoki},
  title   = {Dynamic Ticket Pricing in Sport: An Agenda for Research and Practice},
  journal = {Sport Marketing Quarterly},
  year    = {2012},
  volume  = {21},
  number  = {3},
  pages   = {184--194},
  doi     = {10.1177/106169341202100307}
}

@article{ShapiroDrayer14,
  author  = {Shapiro, Stephen L. and Drayer, Joris},
  title   = {An Examination of Dynamic Ticket Pricing and Secondary Market Price Determinants in Major League Baseball},
  journal = {Sport Management Review},
  year    = {2014},
  volume  = {17},
  number  = {2},
  pages   = {145--159},
  doi     = {10.1016/j.smr.2013.05.002}
}

@article{ShapiroDrayerDwyer16,
  author  = {Shapiro, Stephen L. and Drayer, Joris and Dwyer, Brendan},
  title   = {Examining Consumer Perceptions of Demand-Based Ticket Pricing in Sport},
  journal = {Sport Marketing Quarterly},
  year    = {2016},
  volume  = {25},
  number  = {1},
  pages   = {34--46},
  doi     = {10.1177/106169341602500105}
}

@article{HuangWangHo17,
  author  = {Huang, Yeu Shiang and Wang, Jun Guo and Ho, Jyh Wen},
  title   = {Ticket Pricing with Different Inventory Information Displays},
  journal = {Computers \& Industrial Engineering},
  year    = {2017},
  volume  = {109},
  pages   = {59--70},
  doi     = {10.1016/j.cie.2017.04.035}
}

@article{YangChuWu22,
  author  = {Yang, Yang and Chu, Wan Ling and Wu, Cheng-Hung},
  title   = {Learning Customer Preferences and Dynamic Pricing for Perishable Products},
  journal = {Computers \& Industrial Engineering},
  year    = {2022},
  volume  = {171},
  pages   = {108440},
  doi     = {10.1016/j.cie.2022.108440}
}
\Addresses

\end{document}